\documentclass[12pt,preprint]{emulateapj}
\usepackage{graphicx}

\shorttitle{Stellar Variability of HD~63454}
\shortauthors{Stephen R. Kane et al.}
\slugcomment{Submitted for publication in the Astrophysical Journal}

\begin{document}

\title{Stellar Variability of the Exoplanet Hosting Star HD~63454}

\author{
  Stephen R. Kane\altaffilmark{1},
  Diana Dragomir\altaffilmark{1,2},
  David R. Ciardi\altaffilmark{1},
  Jae-Woo Lee\altaffilmark{3},
  Gaspare Lo Curto\altaffilmark{4},
  Christophe Lovis\altaffilmark{5},
  Dominique Naef\altaffilmark{5},
  Suvrath Mahadevan\altaffilmark{6,7},
  Genady Pilyavsky\altaffilmark{6},
  Stephane Udry\altaffilmark{5},
  Xuesong Wang\altaffilmark{6},
  Jason Wright\altaffilmark{6,7}
}
\email{skane@ipac.caltech.edu}
\altaffiltext{1}{NASA Exoplanet Science Institute, Caltech, MS 100-22,
  770 South Wilson Avenue, Pasadena, CA 91125}
\altaffiltext{2}{Department of Physics \& Astronomy, University of
  British Columbia, Vancouver, BC V6T1Z1, Canada}
\altaffiltext{3}{Department of Astronomy \& Space Science, Sejong
  University, 143-747 Seoul, Korea}
\altaffiltext{4}{ESO, Karl-Schwarzschild Strasse, 2, Garching bei
  M\"unchen, Germany}
\altaffiltext{5}{Observatoire de Gen\`eve, Universit\'e de Gen\`eve,
  51 ch.des Maillettes, 1290 Sauverny, Switzerland}
\altaffiltext{6}{Department of Astronomy and Astrophysics,
  Pennsylvania State University, 525 Davey Laboratory, University
  Park, PA 16802}
\altaffiltext{7}{Center for Exoplanets \& Habitable Worlds,
  Pennsylvania State University, 525 Davey Laboratory, University
  Park, PA 16802}


\begin{abstract}

Of the hundreds of exoplanets discovered using the radial velocity
technique, many are orbiting close to their host stars with periods
less than 10 days. One of these, HD~63454, is a young active K dwarf
which hosts a Jovian planet in a 2.82 day period orbit. The planet has
a 14\% transit probability and a predicted transit depth of
1.2\%. Here we provide a re-analysis of the radial velocity data to
produce an accurate transit ephemeris. We further analyse 8 nights of
time series data to search for stellar activity both intrinsic to the
star and induced by possible interactions of the exoplanet with the
stellar magnetospheres. We establish the photometric stability of the
star at the 3 millimag level despite strong Ca~II emission in the
spectrum. Finally, we rule out photometric signatures of both
star--planet magnetosphere interactions and planetary transit
signatures. From this we are able to place constraints on both the
orbital and physical properties of the planet.

\end{abstract}

\keywords{planetary systems -- techniques: photometric -- techniques:
  radial velocities -- stars: individual (HD~63454)}


\section{Introduction}
\label{introduction}

The number of known exoplanets has now well exceeded 500, revealing a
large diversity in both planetary properties and orbital
characteristics. In the early days of exoplanet discoveries, one of
the first surprises was that of very short-period planets, the
so-called hot Jupiters. Studies have been undertaken which attempt to
find star--planet interactions between these hot Jupiters and their
host stars, and correlations of stellar activity with planetary
emission spectra \citep{knu10} and surface gravities \citep{har10}
have been detected. Evidence has also been found for a general
increase in chromospheric activity of stars which harbor short-period
planets \citep{can11} and surveys have been undertaken which evaluate
such activity in potential planet search targets \citep{arr11}. Most
of these effects are caused by interactions between coronal magnetic
fields and the magnetospheres of the close-in planets
\citep{coh09,lan09}. Searches for observable signatures of such
interactions have been undertaken for HD~189733 \citep{far10} and
CoRoT-6 \citep{lan11} but the evidence has been
inconclusive. \citet{shk08} observed synchronicity of the Ca II H and
K emission for both HD~179949 and $\upsilon$~And with the rotation of
their respective short-period planets, likely due to interactions with
the stellar magnetic fields.

The 2.82~day period planet orbiting HD~63454 (HIP~37284,
TYC~9385-1045-1) was first detected by \citet{mou05} using radial
velocity data obtained with the High-Accuracy Radial-velocity Planet
Searcher (HARPS) mounted on the ESO 3.6m telescope. The host star is a
relatively young ($\sim 1$~Gyr) late-type (K4V) star. The activity
indicators in the discovery data show that the star is active and they
report radial velocity jitter which is attributed to the stellar
activity. Since the star is young, it is predicted to have a
relatively short rotation period of $\sim 20$~days. This star is
extremely southern in declination ($-78\degr$) and so follow-up
observations of the system have been minimal since the planet's
discovery.

Here we present the results of photometrically monitoring HD~63454 as
part of the Transit Ephemeris Refinement and Monitoring Survey (TERMS;
\citet{kan09}). The star was observed over a two week period in order
to extract variability properties of the star. In particular, we are
interested in variability that may correspond with the radial velocity
jitter and/or the influence of the planet on the star. We find no
evidence of such correlations which places limits on the causation of
stellar activity due to the interactions of the planet. We present
additional HARPS data which refines the period and redetermines the
phase of the planet. We subsequently monitored transit windows which
confirms that this planet does not transit the host star. With such a
small orbital period, we use this result to place a lower limit on the
mass of the planet and an upper limit on the radius of the planet.


\section{Keplerian Orbit and Transit Ephemeris}
\label{keplerian}

The original discovery of HD~63454b presented by \citet{mou05}
included 26 radial velocity measurements acquired using the HARPS
instrument and were subject to an additional analysis by
\citet{bab10}. Here we present 8 additional measurements from HARPS
acquired since then which have been used to refine the orbital
parameters and redetermine the phase of this planet. All 34
measurements are shown in Table \ref{rvs}.

\begin{deluxetable}{ccc}
  \tablewidth{0pc}
  \tablecaption{\label{rvs} HARPS Radial Velocities}
  \tablehead{
    \colhead{Date} &
    \colhead{Radial Velocity} &
    \colhead{Uncertainty} \\
    \colhead{(JD -- 2440000)} &
    \colhead{(km\,s$^{-1}$)} &
    \colhead{(km\,s$^{-1}$)}
  }
  \startdata
  13047.625113 &  33.78298  &   0.00321 \\
  13060.629178 &  33.90876  &   0.00435 \\
  13061.601527 &  33.79109  &   0.00308 \\
  13063.592466 &  33.89097  &   0.00290 \\
  13064.637653 &  33.78317  &   0.00366 \\
  13066.590014 &  33.87237  &   0.00271 \\
  13145.500240 &  33.85952  &   0.00268 \\
  13146.479839 &  33.78204  &   0.00240 \\
  13147.463245 &  33.89531  &   0.00551 \\
  13151.455026 &  33.83624  &   0.00323 \\
  13152.452481 &  33.82294  &   0.00345 \\
  13153.473472 &  33.91101  &   0.00199 \\
  13156.443421 &  33.90046  &   0.00316 \\
  13158.466001 &  33.85960  &   0.00485 \\
  13295.880778 &  33.78365  &   0.00177 \\
  13314.841071 &  33.83579  &   0.00161 \\
  13340.808816 &  33.79013  &   0.00173 \\
  13342.779095 &  33.85122  &   0.00232 \\
  13344.787166 &  33.90508  &   0.00175 \\
  13346.788259 &  33.79824  &   0.00140 \\
  13369.736742 &  33.86615  &   0.00142 \\
  13371.762067 &  33.78346  &   0.00186 \\
  13372.728689 &  33.88211  &   0.00178 \\
  13375.769216 &  33.89765  &   0.00194 \\
  13377.779025 &  33.80591  &   0.00321 \\
  13400.742286 &  33.85003  &   0.00193 \\
  13402.640683 &  33.77902  &   0.00309 \\
  13403.736539 &  33.88467  &   0.00157 \\
  13405.746675 &  33.77627  &   0.00181 \\
  13406.654971 &  33.89017  &   0.00164 \\
  13408.668892 &  33.80302  &   0.00173 \\
  13468.516837 &  33.87624  &   0.00232 \\
  15260.618227 &  33.83241  &   0.00095 \\
  15262.516387 &  33.76660  &   0.00077
  \enddata
\end{deluxetable}

The stellar mass according to \citet{mou05} is $M_\star = 0.8 M_\odot$
and the surface gravity is $\log g = 4.23$. More accurate stellar
parameters were measured from HARPS spectra by \citet{sou08} which we
have used to refine the stellar mass and radius. They find the
effective temperature, surface gravity, and metallicity to be
$T_\mathrm{eff} = 4840 \pm 66$~K, $\log g = 4.30 \pm
0.16$~cm~s$^{-2}$, and $[$Fe$/$H$] = 0.06 \pm 0.03$
respectively. Using the polynomial relations of \citet{tor10}, we
derive revised stellar parameters of $M_\star = 0.84 \ M_\odot$ and
$R_\star = 1.05 \ R_\odot$ for HD~63454.

\begin{figure}
  \includegraphics[angle=270,width=8.2cm]{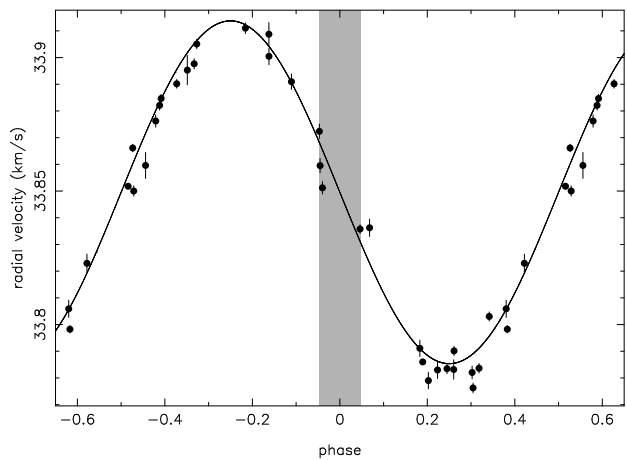}
  \caption{Radial velocity measurements of HD~63454 along with the
    best-fit orbital solution (solid line). The shaded region shows
    the extent of the 1$\sigma$ transit window.}
  \label{rvmodel}
\end{figure}

We fit a single-planet Keplerian solution to the RV data using the
techniques described in \citet{how10} and the partially linearized,
least-squares fitting procedure described in \citet{wri09}. The
inclusion of a linear trend to the solution reduced the
$\chi^2_{\mathrm{red}}$ from 30.68 to 10.14 and the RMS of the
residuals from 10.87 to 6.84 m\,s$^{-1}$. While an offset between the
bulk of the data and the final two measurements of $\sim
20$~m\,s$^{-1}$ would produce comparable reduction in the
$\chi^2_{\mathrm{red}}$, HARPS is known to be extremely stable and
such an offset is not considered plausible. This leads us to favour
the solution which includes a trend. Further radial velocity data is
required to ascertain the precise source of the trend, whether it be
due to the magnetic cycle of the star or the presence of an additional
companion within the system. The adopted solution with the trend is
shown in Table \ref{fitparams}. The parameter uncertainties were
determined from the sampling distribution of each parameter through a
non-parametric bootstrap analysis \citep{fre81}. The folded data and
adopted model with the trend removed are shown in Figure
\ref{rvmodel}.

\begin{deluxetable}{lc}
  \tablecaption{\label{fitparams} Keplerian Fit Parameters}
  \tablehead{
    \colhead{Parameter} &
    \colhead{Value}
  }
  \startdata
  $P$ (days)                       & $2.818049 \pm 0.000071$ \\
  $T_c\,^{a}$ (JD -- 2440000)      & $15583.240 \pm 0.068$   \\
  $T_p\,^{b}$ (JD -- 2440000)      & $13342.870 \pm 0.590$   \\
  $e$                              & $0.000 \pm 0.022$       \\
  $K$ (m\,s$^{-1}$)                & $64.19 \pm 1.65$        \\
  $\omega$ (deg)                   & $87.3 \pm 90.5$        \\
  $dv/dt$ (m\,s$^{-1}$\,yr$^{-1}$) & $-3.95 \pm 0.95$        \\
  $\chi^2_{\mathrm{red}}$          & 10.14                   \\
  RMS (m\,s$^{-1}$)                & 6.84
  \enddata
  \tablenotetext{a}{Time of transit.}
  \tablenotetext{b}{Time of periastron passage.}
\end{deluxetable}




Using the aforementioned stellar mass, we derive a planetary mass of
$M_p \sin i = 0.398 \ M_J$ and a semi-major axis of $a = 0.0368$~AU.
We estimate the planetary radius using the models of \citet{bod03} to
be $R_p = 1.098 \ R_J$. This results in a predicted transit
probability of 14.3\%, a depth of 1.2\%, and a transit duration of
0.13~days. A preliminary search by \citet{mou05} found no evidence for
transits of this planet. In Figure \ref{rvmodel} we include a shaded
region which indicates the calculated size of the transit window
\citep{kan09} at the time of acquiring our photometry. This is
described further in Section \ref{transit} where we present re-phased
photometry which places limits on transits and the inclination or size
of the planet.



\section{Photometry}
\label{photometry}


\subsection{Photometry from Hipparcos}
\label{hipparcos}

\begin{figure}
  \includegraphics[angle=270,width=8.2cm]{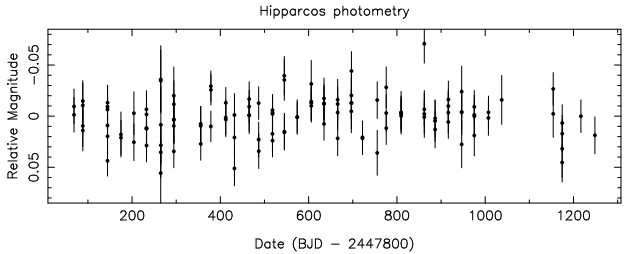}
  \caption{Photometry of HD~63454 from the {\it Hipparcos} mission.}
  \label{phot_hip}
\end{figure}

We investigated the low-frequency photometric stability of HD~63454
using observations from the {\it Hipparcos} satellite. {\it Hipparcos}
observed the star during its three-year mission and acquired a
photometric data set consisting of 124 measurements spanning a period
of 1180 days \citep{per97}, shown in Figure \ref{phot_hip}. The
1$\sigma$ RMS scatter of the 124 HD~63454 measurements is 0.031 mag,
while the mean of the measurement uncertainties is 0.019. The scatter
is roughly 50\% higher than the expected uncertainty of a single
observation, but the range, 0.345 mag, is significantly more than that
expected from a constant star. Consequently, the {\it Hipparcos}
Catalogue described by \citet{per97} lists the variability type for
HD~63454 as a blank, indicating that the star ``could not be
classified as variable or constant.'' We performed a Fourier analysis
of the {\it Hipparcos} data and do not detect any significant periodic
variability. However, this only rules out activity above the 3\%
level. Additionally, the Nyquist frequency of the data is
0.0525~days$^{-1}$ which is slightly above the predicted frequency of
the stellar rotation, thus resulting in substantial aliasing at
smaller periods. The strongest peaks in the periodogram occur at 0.25
and 0.27 days but the power of these peaks are very low.

\subsection{Photometry from CTIO}

Observations of HD~63454 were carried out at the Cerro Tololo
Inter-American Observatory (CTIO) 1.0m telescope using the Y4KCam
Detector\footnote{http://www.astronomy.ohio-state.edu/Y4KCam/}, which
is a 4k$\times$4k CCD with a field of view of about 20 arcminutes on
the side. The target was observed along with three comparison stars
with a Johnson $V$ band filter for 8 nights during the period 22--30
January 2011. An additional night of data was acquired using this
telescope on the night of 5 April 2011 in order to complete phase
coverage of the transit window (see Section \ref{transit}). The
brightness of the target ($V = 9.37$) led to exposure times of 8--12
seconds, high enough to eliminate the effects of shutter errors. The
principal target and comparison stars were carefully placed on
cosmetically clean regions of the CCD and kept in exactly the same
place during the monitoring sequences to avoid inter-pixel
sensitivities.

The target star HD~63454 is known in the 2MASS catalog as
2MASS~J07392187-7816442 \citep{skr06}. There is a nearby fainter star
6.14 arcsecs away (2MASS~J07391989-7816428) for which the $JHK$
magnitudes are 10.636, 12.523, and 12.350 respectively. According to
the photometric quality flags of the 2MASS catalog, the $J$ value
represents an upper limit on the magnitude (i.e., represents a minimum
brightness for the star). The $H-K$ value of 0.173 means the star is
an early M star if on the main sequence.

Aperture photometry was performed on each star by extracting small
regions from the image, $\pm$~100 pixels from the estimated center of
the stellar point-spread function (PSF). The size of the photometric
aperture was limited to restrict light contamination from the nearby
star which was particularly important during nights of bad seeing
which may cause the PSF to spread further into the aperture.  To
achieve sufficient precision to detect low-amplitude variability
including transit signals, we performed relative photometry using the
methods described in \citet{eve01}. The resulting photometry were
binned into equal time intervals of 5 minutes each and are shown in
the top panel of Figure \ref{phot_ctio}. For most nights the 1$\sigma$
RMS was less than 3 millimags, but the combined dataset has a
1$\sigma$ RMS of 3.4 millimags.


\section{Photometric Fourier Analysis}

\begin{figure*}
  \begin{center}
    \begin{tabular}{cc}
      \multicolumn{2}{c}{\includegraphics[angle=270,width=16.8cm]{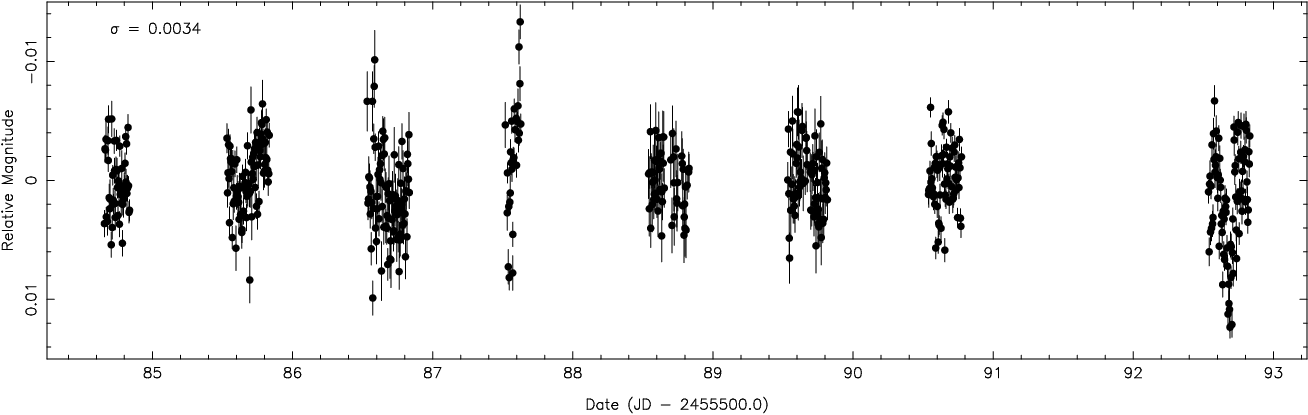}} \\
      \includegraphics[angle=270,width=8.2cm]{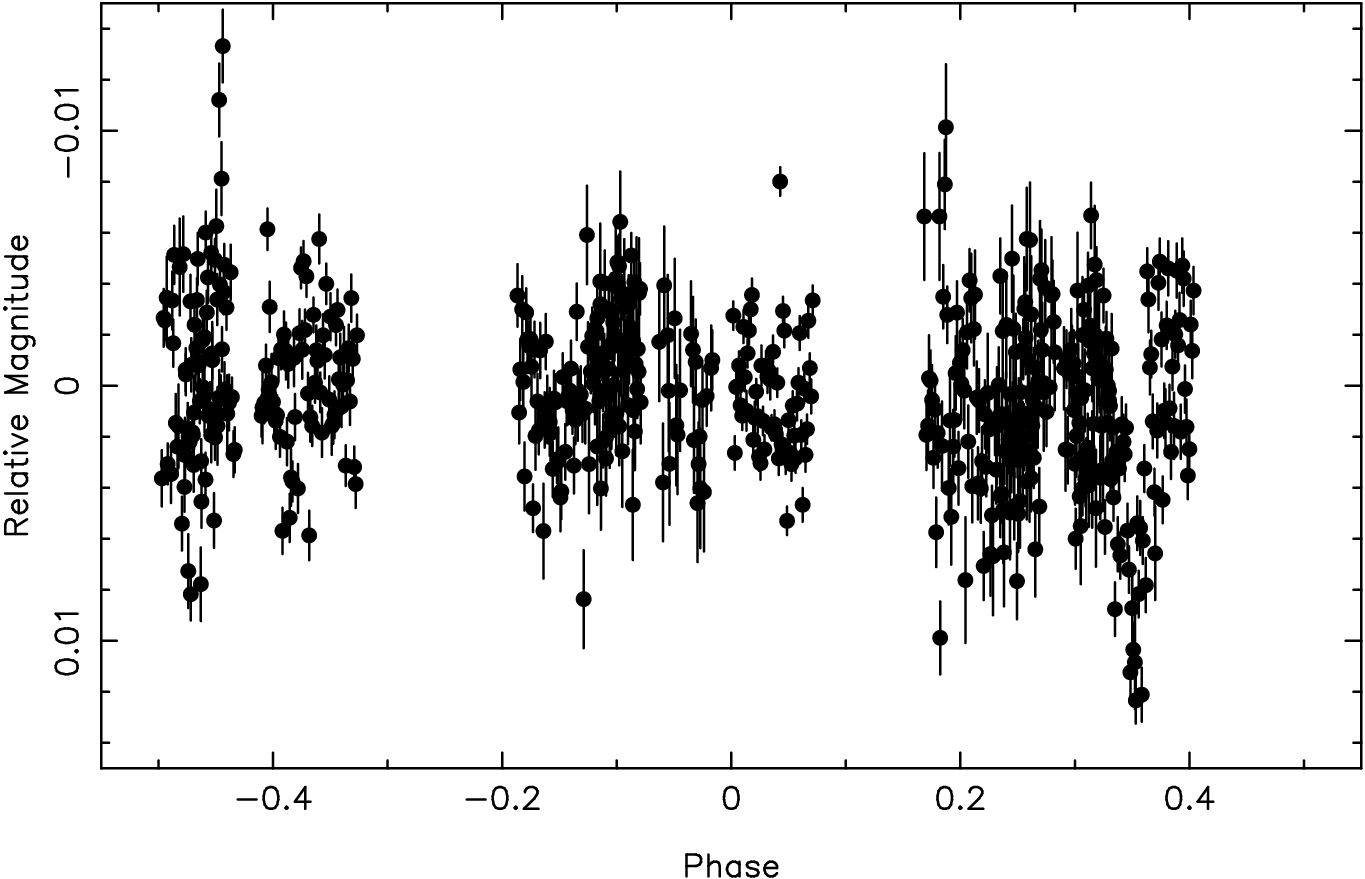} &
      \includegraphics[angle=270,width=8.2cm]{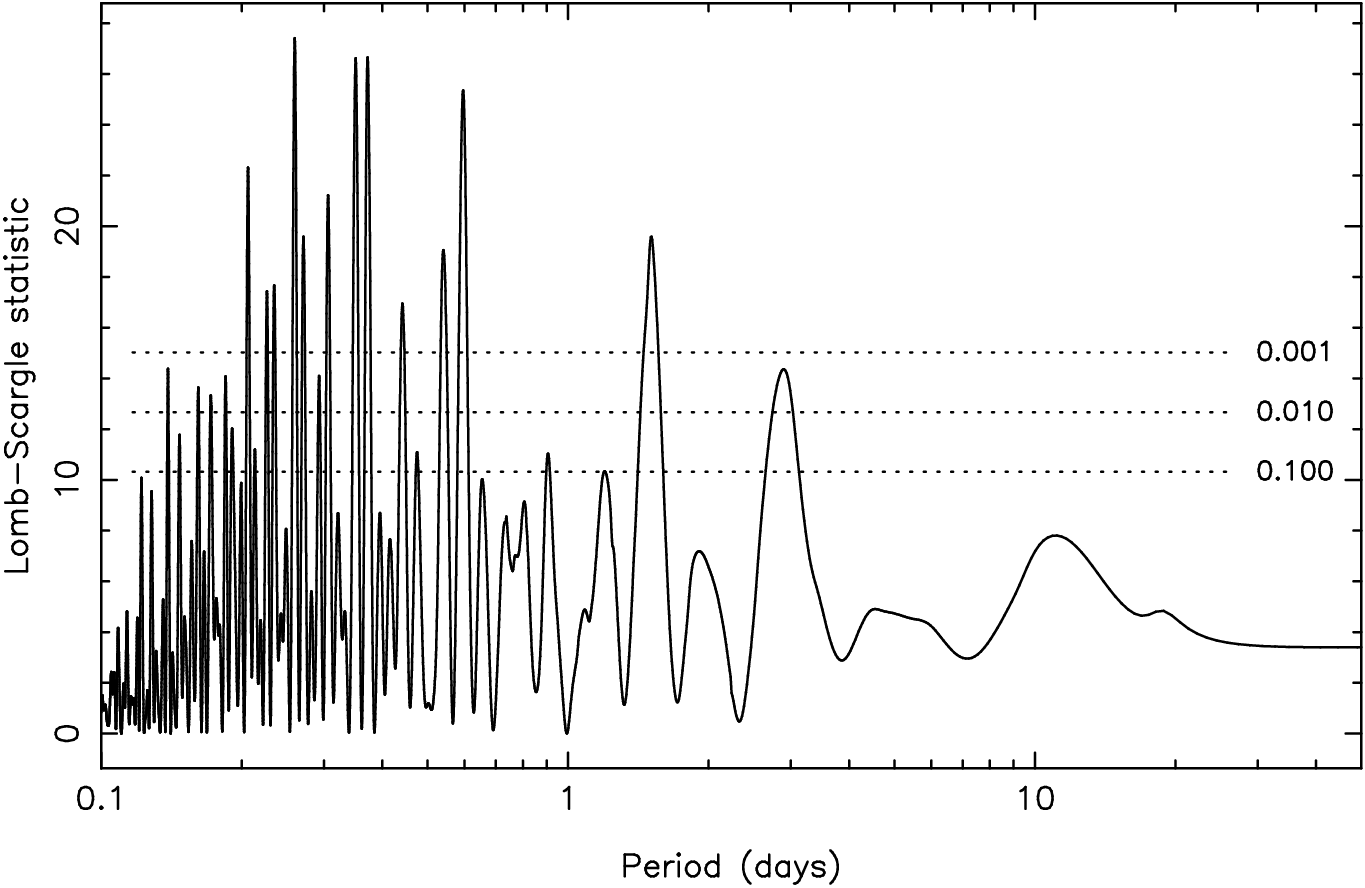} \\
    \end{tabular}
  \end{center}
  \caption{Top panel: Photometry of HD~63454 from the January
    observing run at CTIO, where the data has been binned into 5
    minute intervals. Bottom-left panel: All CTIO photometry folded on
    the best fit period from Table \ref{fitparams} with the predicted
    transit time at phase zero. Bottom-right: Weighted L-S periodogram
    of the Janurary CTIO photometry where dotted lines indicate
    thresholds of false-alarm probabilities.}
  \label{phot_ctio}
\end{figure*}

Here we describe an analysis of the photometry for the purposes of
studying the stability of the star. To investigate the high-frequency
variability of HD~63454, we used a weighted Lomb-Scargle (L-S) fourier
analysis, similar to that described by \citet{kan07}. In particular,
we are interested in activity related to the magnetic cycle and
interactions of the magnetic field and chromosphere with the planet on
the short timescales of its orbital period. Investigation of the line
bisector inverse slope by \citet{mou05} found no correlation with the
orbital period. In the bottom-left panel of Figure \ref{phot_ctio} we
show the complete CTIO dataset folded on the orbital period from Table
\ref{fitparams}. Phase zero in this figure is the location of the
predicted transit time of the planet.

As described by \citet{daw10}, aliases in periodograms result from
discrete sampling times which occur to a lesser degree with unevenly
sampled data. The periodogram of the January 2011 photometry is shown
in the bottom-right panel of Figure \ref{phot_ctio}. There are
significant aliases at periods less than 1 day which are harmonics of
the observing schedule, such as 0.20, 0.26, 0.60, and 1.50 days. Of
note is the strongest peak located at 0.26 days since this lies
between the two strongest peaks located in the Hipparcos data (see
Section \ref{hipparcos}). This is assumed to be the result of the
cadence and resulting Nyquist frequency in each dataset, but we note
it here as a possible indicator of low-amplitude high-frequency
activity. Observations each night lasted $\sim 0.3$~days which results
in broadening of the peaks in the spectral window function and a
double peak at 0.35 and 0.37 days. The strongest feature in the
periodogram beyond a period of 1.5 days is a peak located at
2.90~days. Although tantalizingly close to the measured orbital period
of 2.82~days, the broadness of the peak and strength of the signal are
inconclusive as to a planetary origin, particularly when it is
equally close to an expected alias at 3.00 days. By way of contrast,
the magnetic activity in CoRoT-6 exhibits a photometric variation with
a period of 6.4 days and an amplitude of 2.7\% \citep{lan11}. We thus
conclude that the star is stable at the 3 millimag level over both
stellar rotation and planetary orbital timescales.

For short-period planets, the orbit may be inside the magnetic field
of the host star in which case an electric field is generated by the
interaction of the stellar and planetary magnetic fields
\citep{jar08}. The planet orbiting HD~63454 is at a separation of
7.5~$R_\star$ which, for the relatively young K dwarf ($\sim 1$~Gyr),
may place it inside the stellar magnetosphere given the expected
rotation period. Thus, an additional test would be to attempt the
detection of a time variable radio flux emission from the star--planet
interaction as was performed for HD~189733b by \citet{far10}.


\section{Planetary Transit Exclusion}
\label{transit}

\citet{mou05} state that their photometry ``showed no planetary
transit'' although they do not present the photometry or the precision
of the measurements. Here we present our photometry acquired during
transit windows based upon the revised orbital parameters and
ephemeris and discuss limits on the implied properties of the planet
and orbital inclination. To demonstrate the importance of refining
orbital parameters, the observations during the January CTIO run were
designed to cover two transit windows based upon the orbital
parameters of \citet{mou05}. However, the orbital fit to the updated
HARPS data shifted the predicted windows into the observing gaps in
orbital phase, thus necessitating the additional night of data in
April.

\begin{figure}
  \includegraphics[angle=270,width=8.2cm]{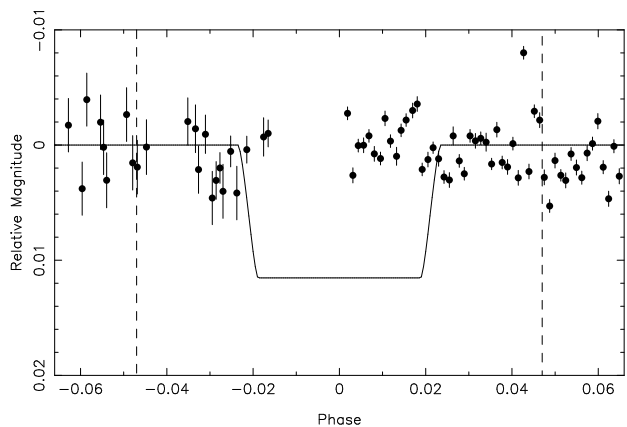}
  \caption{Zoom-in of the CTIO photometry phased on the orbital period
    with the time of mid-transit at phase zero. The solid line is the
    predicted transit signature for the predicted stellar/planetary
    radii and the dashed lines indicate the 1$\sigma$ extent of the
    transit window.}
  \label{tranphot}
\end{figure}

Figure \ref{tranphot} shows a zoomed-in version of the lower-left
panel of Figure \ref{phot_ctio}, where the data once again has been
phased on a zero-point which is the location of the predicted transit
mid-point. The vertical dashed lines indicate the 1$\sigma$ extent of
the transit window which is the predicted duration plus twice the
transit mid-point uncertainty (see Section \ref{keplerian}). In this
case, the transit window size is evenly split between the duration and
mid-point uncertainty yielding a total transit window size of
0.26~days~$=$~0.094 orbital phase. We calculated the predicted transit
signature based upon the analytic models of \citet{man02}, overplotted
as a solid line in the figure.

The conditions on the observing night in April 2011 were exceptional
which produced the photometry that dominates the right-hand side of
Figure \ref{tranphot}. The scatter in these data is larger than what
is expected from photon counting statistics. This excess may be due to
the nearby faint star but is more likely due to stellar photometric
variations. The photometry for that night has a 1$\sigma$ RMS scatter
of 2.3 millimags. The predicted transit depth (1.2\%) is therefore
ruled out at the 5.4$\sigma$ level. This means that, for a
non-transiting planet, the orbital inclination of the planet is
restricted to $i < 81.5\degr$ which results in a lower limit of the
planetary mass of $M_p > 0.402 \ M_J$. On the other hand, if the
planet does transit then the photometric precision rules out planetary
radii of $R_p > 0.77 \ R_J$. A radius just below this threshold would
yield a density of 1.16~g~cm$^{-3}$, resulting in the planet having
remarkably similar properties to the large-cored planet HD~149026b
\citep{sat05}, both in terms of orbital parameters and planetary
characteristics.


\section{Conclusions}

Understanding the star--planet interaction for systems with hot
Jupiters presents an opportunity to further characterize these
planets, particularly for transiting exoplanets where both the mass
and radius of the planet are known. Detection of magnetic field
interactions would yield insight into the internal structure and
rotation rate of these planets. The study of HD~63454 presented here
was conducted as part of the Transit Ephemeris Refinement and
Monitoring Survey (TERMS) in order to detect or rule out both stellar
variability and transit signatures due to the presence of the planet
which has 14\% transit probability and a predicted transit depth of
1.2\%.

This study includes new HARPS radial velocity data in order to
redetermine the phase of the planet during times of photometric
monitoring. The requirement for additional photometry in order to rule
out a transit based upon the revised orbital parameters demonstrates
the need for careful examination of the planetary phase when
monitoring predicted transit windows. The {\it Hipparcos} photometry
reveals no long-term variability of the star, although the sampling
frequency and photometric precision are inadequate to detect
periodicity related to the rotational timescale. The lack of
high-frequency variability at the 3~millimag level may indicate that
the planet is (a) outside the magnetosphere of the star, or (b) the
planetary magnetosphere is very small, or (c) the interaction of the
planetary magnetosphere bow shock with the stellar magnetic field is
best detected at either higher precision or longer wavelengths
(radio). The lack of a transit signature detection indicates that
either the mass of the planet is larger than 0.402~$M_J$ or that the
radius is less than 0.77~$R_J$. In the case of the latter, this would
imply that the planet has very similar properties to HD~149026b.


\section*{Acknowledgements}

The Center for Exoplanets and Habitable Worlds is supported by the
Pennsylvania State University, the Eberly College of Science, and the
Pennsylvania Space Grant Consortium.
J-W.L. acknowledges financial support from the Basic Science
Research Program (grant no. 20100024954) and the Center for Galaxy
Evolution Research through the National Research Foundation of Korea.
This research has made use of
the NASA/IPAC/NExScI Star and Exoplanet Database, which is operated by
the Jet Propulsion Laboratory, California Institute of Technology,
under contract with the National Aeronautics and Space Administration.


\end{document}